\documentclass[%
 reprint,
 amsmath,amssymb,
 aps,
aps,
onecolumn
]{revtex4-2}
\usepackage[utf8]{inputenc}
\usepackage{amsmath}
\usepackage{graphicx}
\usepackage{appendix}
\usepackage{color}
\newcommand{\avg}[1]{\left\langle #1\right\rangle}

\begin{document}
\title{On Random Allocation Models in the Thermodynamic Limit}
\author{Piotr Bialas}\email{piotr.bialas@uj.edu.pl}
\affiliation{Institute of Applied Computer Science, Jagiellonian University,\\ ul. Lojasiewicza 11, 30-348 Krak\'ow, Poland}
\author{Zdzislaw Burda}\email{zdzislaw.burda@agh.edu.pl} 
\affiliation{AGH University of Krakow, Faculty of Physics and Applied Computer Science, \\
al. Mickiewicza 30, 30-059 Krak\'ow, Poland}
\author{Desmond A. Johnston}\email{D.A.Johnston@hw.ac.uk}
\affiliation{School of Mathematical and Computer Sciences, Heriot-Watt University,\\ Riccarton, Edinburgh EH14 4AS, UK }

\begin{abstract}
We discuss the phase transition and critical exponents in the random allocation model (urn model) for different statistical ensembles. We provide a unified presentation of the
statistical properties of the model in the thermodynamic limit, uncover new relationships between the thermodynamic potentials and fill some lacunae in previous results on the singularities of these potentials at the critical point and behaviour in the thermodynamic limit. 

The presentation is intended to be self-contained, so  we carefully derive all formulae step by step throughout. Additionally, we comment on a quasi-probabilistic normalisation of configuration weights which has been considered in some recent studies. 
\end{abstract}

\maketitle

\section{Introduction}
The random allocation model, also known
as the balls-in-boxes model, the urn model or 
the backgammon model \cite{bbj,dgc,bbj2}, is
a simple statistical model describing 
weighted random partitions of particles between boxes.
Despite its simplicity the model 
exhibits very rich critical behavior, 
including discontinuous and continuous phase transitions of different orders depending on the model parameters and the ensemble being considered. The phase transition in the model is related to
a real-space condensation observed in many 
statistical problems including zero-range processes
\cite{s,e,jmp,eh,gl,kmh,g,wbbj},
mass-transport \cite{mez,emz,emz2}, random trees \cite{bb,j}, and
quantum gravity \cite{bbj2,bb2,bbw}. The model 
has been used to understand some aspects of non-equilibrium dynamics of condensate formation \cite{gl}. 
The balls-in-boxes model has also been applied in studies of such diverse problems as wealth condensation \cite{bjjknpz} and the diversity of Zipf's populations \cite{maac}. The model 
can be used to mimic phase separation \cite{klmst,elmm}, 
condensation in complex networks \cite{ahe,bbw2},
fire-ball formation at the van Hove singularity \cite{vH},
formation of the giant component/cluster in networks or percolation models
and the statistics of the longest interval in tied-down renewal
processes \cite{w,g1,g2}. The latter phenomenon is closely related to the appearance of ``big jumps'' in random walks with sub-exponentially distributed jump sizes, as described in \cite{seva}.

In the present paper we revisit the issue of the phase transition in the model,
and analyse it carefully from the point of view of equilibrium statistical mechanics. 
We describe in detail how the order of the transition depends on the parameters of the model in various ensembles and present some new results on the singularities of the thermodynamic potentials and the relationships between them. We also revisit the
issue of finite size effects by illustrating a 
typical evolution of the particle distribution with the increasing 
system size in the condensed phase, which reveals a non-uniform convergence to the limiting distribution. We discuss the interpretation of the deviations of the particle distribution for finite systems
from the limiting distribution in the light of various rigorous results on the nature of the condensate in \cite{gl,gss,fls,cg1,g3,cg2,jcg,agl}. Related work on the nature of the condensate in mass transport models  and finite-size corrections may be found in \cite{mez1, mez2}.
In addition, we discuss some subtleties in  a {quasi}-probabilistic normalisation used in \cite{cg1,maac} which regularizes an otherwise divergent sum over the weights and elucidate its exact relation to the class of models here. 

The paper is organized as follows. In section \ref{sec:canonical}
we introduce a canonical ensemble and the basic quantities 
that describe the behaviour of the system in this canonical ensemble.
In section \ref{sec:thermolimit} we apply the saddle point method to calculate the 
free energy density of the system in the canonical ensemble for low particle densities that are smaller
than some critical density $\rho_{\rm c}$ in a thermodynamic limit in which the number of particles $S$ and boxes $N$ are sent to infinity at some fixed density. A detailed saddle-point calculation of the particle distribution can be found in Appendix \ref{app:piq}. For densities larger than $\rho_{\rm c}$  the system exhibits a real-space condensation. 
In section \ref{sec_phase_t} we note that whether there is a phase transition at a finite critical density or not depends on the asymptotic behaviour of the weights governing the particle distribution.
This divides the weight functions into three families
for which the system has only the fluid phase,  
only the condensed phase, or has both phases 
with a phase transition between them at a finite critical density. 
In section \ref{sec:power-law} we then focus on power-law weights of the form $w(s)=s^{-\beta}$ for $s$ particles in a box to analyse the types of possible phase transitions. The asymptotic properties of the polylogarithm, which is the generating function for power law weights, are recalled in Appendix \ref{app:polylog}. These are then applied in detailed calculations in Appendix \ref{app:s_fenergy}, where  we 
derive the possible scenarios that depend on the power in the weight function and determine the free energy and its behaviour at the critical point by evaluating the asymptotic behaviour of the cumulant generating function of the weights. 
We find that the phase transition is second order for 
$\beta\in (3,+\infty)$. For $\beta=3$, the phase transition is also 
second order but with a logarithmic discontinuity and
for $\beta<3$
the order of the transition increases as $\beta$ approaches $2$, 
eventually disappearing at $\beta=2$.

In section \ref{sec:GC} we repeat these steps for a grand-canonical ensemble in which the number of boxes  is allowed to fluctuate while taking the thermodynamic limit. This reveals that, while the canonical system has a {continuous} phase transition of arbitrary order as $\beta$ was varied, the grand-canonical system
may additionally display  a {\it discontinuous} (i.e. first order) phase transition. The details of these calculations may be found in Appendix \ref{app:s_grand_pot}, where we employ similar asymptotic methods to those used in investigating the canonical ensemble to show that the phase transition in the grand canonical ensemble is  first order for $\beta\in (2,+\infty)$ and that
for $\beta\in (1,2]$ the order of the phase transition varies 
from second to infinite.
It should be noted that the term ``grand-canonical'' is usually used to
refer to statistical systems with a fluctuating number of particles, 
while here we apply it to a system with a fluctuating number of {\it boxes}, which plays the role of the volume of the system. 

The standard grand-canonical ensemble with a fluctuating number of 
particles is discussed in section \ref{sec:oe}. As we will
see, in this case the partition function entirely factorises
so the system is in a sense trivial. An ensemble with a varying number of both particles and boxes is also defined. 
We  move on in section \ref{sec:quasi} to discuss ensembles with the quasi-probabilistic weights of  \cite{cg1,maac} and their relation to the other ensembles discussed here. We conclude with 
a short summary in section \ref{sec:summary} which re-iterates the main results on the phase structure in the canonical and grand-canonical ensembles and emphasizes the Legendre-Fenchel transform relationship between the thermodynamic potentials in the two ensembles. We also note that the grand-canonical potential is just the inverse function of the cumulant generating function of the weights. Finally, we advertise further work using the methods deployed in this paper to evaluate R\'enyi  entropies for zeta-urns and to calculate the partition function zeros of the model.

\section{Canonical ensemble}
\label{sec:canonical}
The balls-in-boxes model is defined by the partition function 
\cite{bbj}
\begin{equation}
    Z_{S,N} = \sum_{(s_1,\ldots,s_N)} 
    w(s_1) \ldots w(s_N) \delta_{S- (s_1+\ldots +s_N)}
    \label{ZSN}
\end{equation}
that describes weighted distributions of $S$ particles in $N$ boxes, where $s_i$'s denote the occupations of boxes $i=1,\ldots,N$. The lowercase $\delta$ 
represents the Kronecker delta: $\delta_n=1$ for $n=0$ and 
$\delta_n=0$ for all other integers $n\ne 0$. The delta 
in Eq. (\ref{ZSN}) selects configurations that have exactly $S$ particles in $N$ boxes.
The statistical weight of a configuration 
$(s_1,s_2,\ldots,s_N)$, that describes the partition of particles
between boxes, is the product of statistical weights $w(s_i)$ of individual boxes, that depend only on the number of particles
in the box. The numbers of particles in different boxes are {\em almost} independent of each other, but 
the constraint $s_1+\ldots+s_N=S$ makes them weakly dependent.
In some conditions, the presence of this constraint leads to a phase transition, as will be seen later. 

In general, the statistical weight $s \rightarrow w(s)$ is a non-negative real-valued function defined for $s=0,1,2,\ldots$. However, we find it convenient to limit the range of $s$ 
to positive integers $s=1,2,\ldots$, in which case each box must contain at least one particle and there 
must be at least $S=N$ particles in a system with $N$ boxes. The
full range (including empty boxes) can be always recovered by introducing new weights 
$w'(s)=w(s+1)$ for $s=0,1,2,\ldots$ and redefining the box occupation numbers
$s'_i = s_i-1$ and $S'=S-N$. In the sequel we assume that there are no empty boxes.
This version of the model occurs naturally in many of the problems mentioned in the introduction.

Let us consider a system with $q$ balls in the first box. Then the rest of the boxes contains $S-q$ balls and are described by the partition function $Z_{S-q,N-1}$. This leads to the recurrence relation
\begin{equation}
    Z_{S,N}=\sum_{q=1}^{S-N+1}w(q) Z_{S-q,N-1}
\label{eq:Z-recurrence}    
\end{equation}
for $S\ge N\geq 1$, and
\begin{equation}
Z_{S,0} =\delta_S 
\label{ZS0}
\end{equation}
for $N=0$. The formula for $Z_{S,0}$ ensures that 
\begin{equation*}
    Z_{S,1}=w(S)
\end{equation*}
as required. 

Consider a physical quantity $O=O(s_1,\ldots,s_N)$. The ensemble average is defined as
\begin{widetext}
\begin{equation}
    \langle O \rangle_{S,N} = \frac{1}{Z_{S,N}} \sum_{(s_1,\ldots,s_N)} O(s_1,\ldots,s_N) w(s_1) \ldots w(s_N) \delta_{S - (s_1+\ldots +s_N)} .
    \label{cO}
\end{equation}
\end{widetext}
It is worth noting that the following transformation
\begin{equation}
 w(s) \rightarrow e^{\lambda + \sigma s} w(s).
    \label{wtransf}
\end{equation}
leaves the ensemble averages (\ref{cO}) unchanged 
\begin{equation}
    \langle O \rangle_{S,N} \rightarrow \langle O \rangle_{S,N}
    \label{OO}
\end{equation}
because it introduces the same factors $e^{\lambda N + \sigma S}$  
in the numerator and denominator of (\ref{cO}) and they cancel out.
We will use this invariance in the sequel many times.

As an example, consider the fraction of sites with $q$ particles in a given configuration $(s_1,s_2,\ldots,s_N)$:
\begin{equation}
\pi(q) = \frac{1}{N} \sum_{i=1}^N \delta_{q-s_i} . 
\label{pi_definition}
\end{equation}
The ensemble average of $\pi(q)$ 
can be calculated directly from the partition function by 
observing that if a box contains $q$ particles then the remaining $N-1$ boxes contain $S-q$ particles.
This leads to the relation
\begin{equation}
    \langle \pi(q) \rangle_{S,N} = \frac{w(q) Z_{S-q,N-1}}{Z_{S,N}}
    \label{piq}
\end{equation} 
for $S\ge N\geq 1$ and $q=1,2,\ldots, S-N+1$. Relation \eqref{eq:Z-recurrence} ensures that this distribution is normalised. 
\begin{equation}
    \sum_{q=1}^{S-N+1} \langle \pi(q) \rangle_{S,N} = 1 .
\end{equation}

As an illustration consider the model with weights $w(s)=1$ for $s=1,2,\ldots$.
One finds that 
\begin{equation}
Z_{S,N} = \binom{S-1}{N-1} ,
\label{Ztrivial}
\end{equation}
and
\begin{equation}
    \langle \pi(q) \rangle_{S,N} = \frac{\binom{S-1-q}{N-2}}{\binom{S-1}{N-1}} 
    \label{pi1}
\end{equation}
for $q=1,2,\ldots$. Note that the invariance of the ensemble
averages (\ref{OO}) under the transformation (\ref{wtransf})
leads to the somewhat counter-intuitive conclusion 
that exponentially increasing or decreasing weights $w(s)=e^{\lambda+\sigma s}$ give exactly the same 
results as the constant weights $w(s)=1$. In
particular, the particle distribution for the exponentially
increasing or decreasing weights will be given by (\ref{pi1}).
This is directly related to the well-known
probabilistic fact that exponential random
variables, conditioned on their sum, are uniformly distributed.

\section{Thermodynamic limit} 
\label{sec:thermolimit}

We are interested in the behaviour of the system in the thermodynamic limit:
\begin{equation}
   S\rightarrow \infty , \  \frac{N}{S} \rightarrow r  
   \label{dlimit}
\end{equation}
where $r \in (0,1)$. The behaviour depends on the form of $w(s)$ but also on $r$ which is a free parameter of the model, being 
the reciprocal of the average particle density $\rho$:
\begin{equation}
    \rho\equiv\frac{S}{N} =\frac{1}{r}.
\end{equation}
We will use $r$ and $\rho$ interchangeably.

To describe the behaviour of the system in the thermodynamic limit (\ref{dlimit}), it is convenient to introduce a thermodynamic potential
\begin{equation}
 \phi(r) = \lim \frac{1}{S} \log Z_{S,N} 
\label{phi}
\end{equation}
where ``lim'' in this equation means the limit (\ref{dlimit}). With a mild misuse of terminology $\phi(r)$ may be called the free energy density
(free energy per particle).
The free energy density, $\phi(r)$, is the rate of the asymptotic growth of the partition function with the system size for given $r$ 
\begin{equation}
   Z_{S,N} \propto e^{S \phi(r)}
   \label{Zphi}
\end{equation}
in the thermodynamic limit (\ref{dlimit}). The sub-exponential corrections are omitted. For trivial weights, that is for $w(s)=1$ for $s=1,2,\ldots$, discussed in the previous section, the partition function (\ref{Ztrivial}) behaves in this limit asymptotically as
\begin{equation}
    Z_{S,N} \propto \frac{1}{\sqrt{2\pi S r(1-r)}} 
    e^{-S[r\log r + (1-r) \log(1-r)]}
    \label{Za}
\end{equation}
so the free energy density is 
\begin{equation}
    \phi(r) =  -r\log r - (1-r) \log(1-r) .
\end{equation}
This can be seen by applying Stirling's formula to (\ref{Ztrivial}).
It is also easy to see that the particle distribution 
(\ref{pi1}) takes the following asymptotic form 
in the limit (\ref{dlimit})
\begin{equation}
    \langle \pi(q) \rangle_{S,N} \propto \frac{r}{1-r} e^{-q \log \frac{1}{1-r}}
    \label{pia}
\end{equation}
for $q=1,2,\ldots$. The asymptotic expressions
(\ref{Za}) and (\ref{pia}) hold for any $r\in (0,1)$.
They break down for $r=0$, (infinite density) that is when the number 
of particles $S$ grows faster than linearly with the number 
of boxes $N$ as $N$ goes to infinity. They also break down
for $r=1$, that is when the difference $S-N$ grows 
slower than linearly with $N$ as $N$ goes to infinity. 

In the general case, the thermodynamic potential $\phi(r)$
can be calculated using the saddle point  method.
Using an integral representation of the Kronecker delta
\begin{equation}
   \delta_n = \int_{-\pi}^{\pi} \frac{d\alpha}{2\pi} e^{i n \alpha} .
   \label{delta_integral}
\end{equation}
we can rewrite the partition function (\ref{ZSN}) 
as follows
\begin{equation}
    Z_{S,N} = \int_{-\pi}^{\pi} \frac{d\alpha}{2\pi} e^{i S \alpha}
    \left( \sum_{s=1}^\infty w(s) e^{-i s \alpha} \right)^N .
    \label{integral_ZSN}
\end{equation}
The right hand side of this equation 
can be more concisely written as 
\begin{equation}
    Z_{S,N} = \int_{-\pi i}^{\pi i} \frac{d\alpha}{2\pi i} e^{S \alpha +  N K(\alpha)}
    \label{rotated}
\end{equation}
where $K(\alpha)$ is a cumulant generating function
\begin{equation}
   K(\alpha) = \log \sum_{s=1}^\infty w(s) e^{- s \alpha} .
   \label{K}
\end{equation}
In the limit (\ref{dlimit}), the leading 
contribution to the integral is
\begin{equation}
    Z_{S,N} \propto e^{S\phi(r)}
    \label{spZ}
\end{equation}
where 
\begin{equation}
    \phi(r) = \alpha(r) + r K(\alpha(r)).
    \label{sp_phi}
\end{equation}
and $\alpha(r)$ 
is a solution of the saddle point equation.
\begin{equation}
    \frac{1}{r} = - K'(\alpha(r)) .
   \label{alphastar}
\end{equation}
This result is obtained by deforming the integration contour 
so that it passes through a saddle point.
The saddle point is located on the real axis and its
position $\alpha(r)$ depends on the particle density.
When the particle density $\rho=1/r$ increases, $\alpha(r)$
decreases. The minimal value that it may take is limited
by $\alpha_{\rm c} = -\log R_{\rm c}$, where
\begin{equation}
R_{\rm c} = \limsup_{s\rightarrow \infty} \frac{1}{ w(s)^{1/s}}
\end{equation}
is the radius of convergence of the series (\ref{K}). For the monotonic weights that we 
consider here
\begin{equation}
\alpha_{\rm c} = \lim_{s\rightarrow \infty} \,  \frac{\log w(s)}{s} .
\label{alpha_min}
\end{equation}
For $\alpha > \alpha_{\rm c}$ the series in (\ref{K}) is convergent. The saddle point solution $\alpha(r)$ holds
only for $r>r_{\rm c}$ 
\begin{equation}
    r_{\rm c} = - \frac{1}{K'(\alpha_{\rm c})} .
    \label{rcr}
\end{equation}
For $r\rightarrow r_{\rm c}^+$, the saddle point 
$\alpha(r) \rightarrow \alpha_{\rm c}$ approaches the singularity 
of $K(\alpha)$ (\ref{K}).

In the thermodynamic limit (\ref{dlimit}), one can also derive (see Appendix \ref{app:piq}) an asymptotic form of the particle distribution (\ref{piq})
\begin{equation}
  \langle \pi(q) \rangle_{S,N} \propto \frac{w(q) e^{-q\alpha(r)}}{e^{ K(\alpha(r))}} =
   \frac{w(q) e^{-q \alpha(r)}}{\sum_{s=1}^\infty 
   w(s) e^{-s \alpha(r)}} .  
   \label{exp_sup}
\end{equation}
We see that the distribution is suppressed by an exponential factor $\sim \exp\left( -q(\alpha(r)-\alpha_{\rm c}) \right)$ 
for large $q$.

The first moment of the particle distribution
should be exactly equal to the particle density
\begin{equation}
    \frac{1}{r} = \langle q \rangle_{S,N} \equiv  
    \sum_{q=1}^\infty q\cdot \langle \pi(q) \rangle_{S,N} .
    \label{q1}
\end{equation}
Replacing  $\langle \pi(q) \rangle_{S,N}$ on the right hand side of this equation with (\ref{exp_sup})
we find
\begin{equation}
    \frac{1}{r} = 
    \frac{\sum_{q=1}^\infty q \cdot w(q) e^{-q\alpha(r)}}{\sum_{s=1}^\infty w(s) e^{-s\alpha(r)}} = -K'(\alpha(r)) 
    \label{q2}
\end{equation}
consistently with the saddle point equation (\ref{alphastar}). 

  Above $\rho_{\rm c}$ another {\em condensed} phase appears - the condensed phase where an extensive (proportional to $S$) number of particles condense in one box.
This was first observed in \cite{bbj} and then rigorously proven in \cite{gss,fls,cg1,agl,cg2,jcg}. In the next section we discuss in detail the phase transition between the two phases. 

\section{Phase structure} 
\label{sec_phase_t}

As mentioned, the system can have two distinct phases:
the fluid phase for small particle densities and the condensed phase 
for large particle densities.
The fluid phase is described by the solution of the saddle point equations \eqref{alphastar} which correspond to a local solution
of the maximisation problem within the interval $(r_{\rm c},1)$. In the condensed phase, the equations are no longer valid. In this case
the maximal contribution to the free energy comes from the boundary at $r_{\rm c}$.

The critical density at which the transition from one phase to another occurs is given by $\rho_{\rm c}=1/r_{\rm c}$ with $r_{\rm c}$ defined by \eqref{rcr}. Whether the system has a phase transition or not depends on the weight function $w(s)$. Let us discuss the possible scenarios.

The set of possible weight functions can be divided 
into three families which are classified by the value of the parameter $\alpha_{\rm c}$ (\ref{alpha_min}): the first family includes weight functions $w(s)$ for which $\alpha_{\rm c}=-\infty$, the second includes weights for which $\alpha_{\rm c}=+\infty$, and the third weights for which $\alpha_{\rm c}$ is a finite number.

\begin{itemize}
\item
Weight functions $w(s)$ from the first family ($\alpha_{\rm c}=-\infty$) fall off asymptotically to zero faster than exponentially for $s\rightarrow \infty$, for instance 
$w(s) = 1/s!$, $w(s) = e^{-s^2}$. Weights which vanish
for all $s$ above a certain value $s_0$: $w(s)=0$ for all $s>s_0$ also belong to this category.

\item Weight functions $w(s)$ from the second family
($\alpha_{\rm c}=+\infty$) grow faster 
than exponentially for $s\rightarrow \infty$, 
for instance $w(s) = s!$ or $w(s)=e^{s^2}$. 

\item Weight functions $w(s)$ from the third family
($-\infty<\alpha_{\rm c}<+\infty$) neither increase nor 
decrease faster than exponentially for $s\rightarrow \infty$, 
for instance $w(s)= e^{\sigma s} s^{-\beta}$ or $w(s) = e^{\pm \sqrt{s}}$.
One should note that power-like weights, $w(s)=s^{-\beta}$, 
belong to this category, even if $\beta$ is negative.

\end{itemize}

For the first family the system has only the fluid phase,
for the second one it has only the condensed phase. 
The third family, in a way, interpolates between the first two cases
and therefore it is the most interesting one. 

From now on, we focus on 
weight functions from the third family, that is 
such that $\alpha_{\rm c}$ (\ref{alpha_min}) is finite.
Using the transformation (\ref{wtransf}), which does not affect the ensemble averages, we can transform the weights $w(s) \rightarrow e^{-\alpha_{\rm c} s} w(s)$ 
so that the critical value (\ref{alpha_min}) after the transformation 
is $\alpha_{\rm c}=0$. So from now on, unless we specify otherwise, we set by default $\alpha_{\rm c}=0$.

The inverse particle density $r$ is a free parameter that can change in the range 
$r \in (0,1)$. The saddle point equation \eqref{alphastar} holds for $r \in (r_{\rm c}, 1)$ where $r_{\rm c}$ is given by \eqref{rcr} (with $\alpha_{\rm c}=0)$. Two situations can 
be distinguished: $-K'(0)$ is infinite or finite. In the former case the critical density 
is infinite, so there is no phase transition and
the saddle point solution holds for the whole range of $r \in (0,1)$. 
In the latter case, the critical density is finite, so there is a phase transition
at $\rho_{\rm c} = -K'(0)$. The saddle point solution holds for $r \in (r_{\rm c},1)$. 
As mentioned, in this case the particle distribution (\ref{exp_sup}) falls off exponentially 
for large $q$ as $\pi(q) \propto w(q)e^{-q \alpha(r)}$ with $\alpha(r)>0$. 
At $r=r_{\rm c}$ the exponential factor disappears, $\alpha(r_{\rm c})=0$, and the particle
distribution (\ref{exp_sup}) tends in the thermodynamic limit (\ref{dlimit}) to its critical form
\begin{equation}
    \langle \pi(q) \rangle_{S,N} \rightarrow \pi_{\rm c}(q) \equiv 
    \frac{w(q)}{K(0)} = \frac{w(q)}{\sum_{s=1}^\infty w(s)}
    \label{pic}
\end{equation}
whose mean is equal to the critical density 
\begin{equation}
\sum_{q=1}^\infty q \pi_{\rm c}(q) = -K'(0) = \rho_{\rm c}.
\end{equation} 
To see what happens for $\rho>\rho_{\rm c}$, in the condensed 
phase, we perform a finite size analysis.
More precisely, we will numerically determine the particle distribution $\avg{\pi(q)}_{S,N}$ for finite $N$ using formula \eqref{piq} and the recursion \eqref{eq:Z-recurrence}.
In fact, instead of (\ref{eq:Z-recurrence}) 
we have used the following recursion 
\begin{equation}
    Z_{S,2N} = \sum_{q=N}^{S-N} Z_{q,N}Z_{S-q,N}
\end{equation}
in the calculations. It is more efficient than 
\eqref{eq:Z-recurrence}, because it doubles $N$ in one step, 
while \eqref{eq:Z-recurrence} increases $N$ by one. 

\begin{figure}
    \centering
    \includegraphics[width=0.44\textwidth]{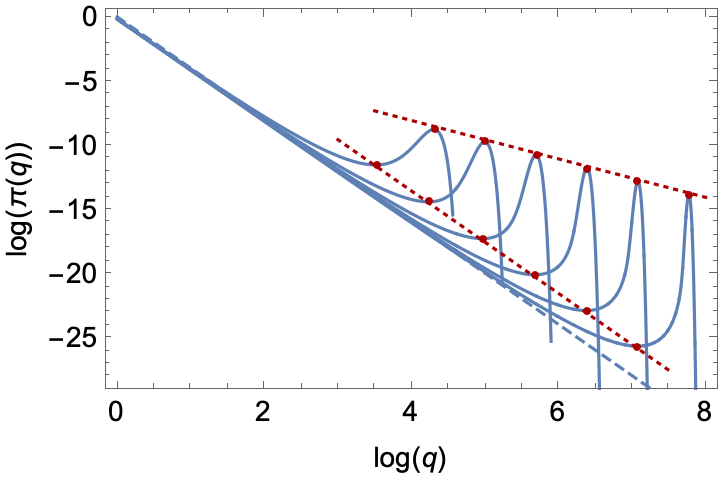} \qquad
    \includegraphics[width=0.44\textwidth]{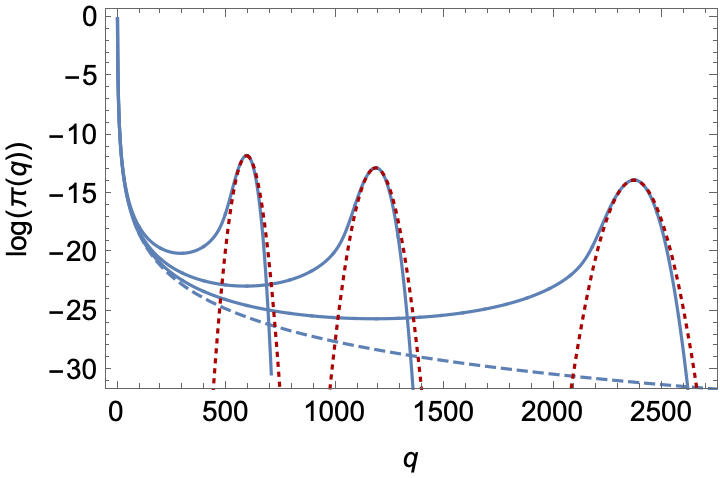}
    \caption{The distribution $\langle \pi(q) \rangle_{S,N}$ for $N=2^k+1$ for $\beta=4$ and $S = \lceil 1.4 N \rceil$.
    Left: Plots of the distribution for $k=8,9,\ldots 13$ in logarithmic scale. 
    The blue dashed line represents the critical distribution $\pi_{\rm c}(q) = q^{-4}/\zeta(4)$. 
    The red dashed lines pass through the local maxima and minima of the graphs. 
    The heights of the maxima scale as $N^{-3/2}$ (\ref{Gp}) 
    and of the minima as $N^{-4}$ \cite{gl}. Right: Plots of the distribution for $k=11,12,13$ in semi-logarithmic scale. As before, the blue dashed line represents the critical
    distribution. The red dotted lines represent Gaussian curves with $S - (N-1) K'(0)$ means,
    $(N-1)K''(0)$ variances, and $1/N$ normalisation.
    }
    \label{fig:piq}
\end{figure}

As an example, we study the behaviour of the system for
power-law weights
\begin{equation}
    w(s)=s^{-\beta}
\end{equation}
with $\beta=4.0$ and $\rho=1.4$, which is 
greater than the critical density 
\begin{equation}
    \rho_{\rm c}=\frac{\zeta(3)}{\zeta(4)}\approx 1.11,
\end{equation}
so the system is in the condensed phase.
The results are presented in  Fig.~\ref{fig:piq}. 
As we can see from the first picture the distribution can be divided into a ``bulk'' part corresponding to the critical distribution
\begin{equation}
    \pi_{\rm c}(q)=\frac{q^{-4}}{\zeta(4)}
    \label{pi4}
\end{equation}
denoted by the dashed line in the figure and a peak.  
The peak is located at $Q \approx N(\rho-\rho_{\rm c})$.
The area under the peak tends to $1/N$. 
This can be  clearly seen in Fig.\ref{fig:piq} (right) where the peaks
are compared to Gaussian curves with area $1/N$.
This picture can be understood as a split of the system 
into a critical subsystem consisting of $N-1$ boxes and a single box that captures the excess particles that did not fit in the critical subsystem. 
This picture was conjectured in \cite{bbj} and then rigorously proven, see for instance \cite{gl,gss,fls,cg1,g3,cg2,jcg,agl}.

The critical part consists of $S_{N-1}$ 
particles in $N-1$ boxes. 
$S_{N-1}$ can be approximated as a sum of $N-1$ independent random numbers $q_1+\ldots + q_{N-1}$
distributed according to the critical distribution $\pi_{\rm c}(q)$. By the generalised central 
limit theorem \cite{gk}, such a sum grows on average as $(N-1) \rho_{\rm c}$, and behaves in the limit
$N\rightarrow \infty$ like a normal random variable with the variance $(N-1) K''(0)$ if $K''(0)<\infty$, 
or as a one-sided $\alpha$-stable random variable with $\alpha \in (1,2)$, otherwise. This tells us
that the number of particles in the remaining box is on average $Q=S - S_{N-1} \approx N(\rho-\rho_{\rm c})$,
and its fluctuations about the mean are of order $\sqrt{N}$ if $K''(0)<\infty$ or $N^{1/\alpha}$ otherwise. 

In the discussed example $K''(0)<\infty$, so the peak 
can be approximated by a normalised Gaussian curve with mean $Q$ and variance $N K''(0)$
\begin{equation}
\frac{1}{N} \pi_{\rm cond}(q) \approx \frac{1}{N} 
\frac{1}{\sqrt{2 \pi N K''(0)}} \exp -\frac{(q-Q)^2}{2 N K''(0)} 
\label{Gp}
\end{equation}
with an additional $1/N$ factor that reflects the fact that the condensate is in one of the $N$ boxes.  
The height of the peak is of order $N^{-3/2}$ as follows
from the last equation. 

The approach to the limit distribution (\ref{pi4})
is very slow and nonuniform as shown in Fig. \ref{fig:piq}. 
The peak which is present for any finite $N$
does not disappear but moves away to infinity when $N$ increases. 
The shape of the peak deviates from the Gauss-curve shape. The deviations are seen as long tails on the left side of parabolas in a semi-logarithmic plot. Slight deviations of tails on the right
hand side can also been seen. The left tails
are remnants of the power-law tail $q^{-\beta}$ of the critical distribution (\ref{pic}) \cite{gl}. The contribution
from the tails decreases as $N$ grows, as discussed below. 
The excess probability accumulated in the left tails
\begin{equation}
    P_{\rm dev} = \sum_{q=1}^{q_{\max}} \left( 
    \langle \pi(q) \rangle_{S,N} - \frac{N-1}{N} \pi_{\rm c}(q) - \frac{1}{N}
    \pi_{\rm cond}(q) \right)
    \label{eq:Pdev}
\end{equation}
decreases as $N$ increases. In (\ref{eq:Pdev}) $q_{\max}$ is 
the position of the peak maximum.
The excess probability is calculated as the area between the curve $\langle \pi(q) \rangle_{S,N}$ and a curve obtained as a weighted sum of the limiting distribution (\ref{pi4}) and the Gaussian peak (\ref{Gp}), to the left of the maximum of the peak. The weights $(N-1)/N$ and $1/N$ correspond to the fractions of boxes occupied by the critical subsystem 
and the condensate, respectively. 

For example, for the three graphs shown in the right plot 
in Fig. \ref{fig:phi}, which correspond to $N=2^k+1$ with $k=11,12,13$, the excess probability accumulated in the tail as compared to the probability $1/N$
accumulated in the peak, $P_{\rm dev}/(1/N)$, is approximately 
$4.2\%$, $2.8\%$ and $1.8\%$, respectively. 
To summarise, the analysis shows that there is a single peak 
in the particle distribution that detaches from the bulk.
The area under this peak is approximately $1/N$,
which means that the peak comes from a single box, where the
condensate resides. The bulk part of the distribution approaches the critical distribution with deviations from the limiting shape
being finite size effects. This picture was rigorously proven in \cite{gl,gss,fls,cg1,g3,cg2,jcg,agl}. 
The details of the approach to the thermodynamic limit determines the origin of the sub-leading corrections.
As noted in the introduction \cite{s,e,jmp,eh,gl,kmh,g,wbbj} 
 the particle distribution can be regarded as being the non-equilibrium steady state of a zero range process with suitably chosen rates.
 The relocation dynamics of the condensate at the critical density will then, with suitable scaling, be dominated by switching to a (distributed) meta-stable fluid \cite{agl,cg2,jcg}. On the other hand, if the density is fixed to be strictly larger than the critical density in taking the thermodynamic limit the condensate will transfer via a sub-condensate   and distributed fluid configurations will eventually be much less likely than sharing the excess mass between just two sites \cite{gl,g3}.
The sub-condensate is responsible for the shape of the deviations in the elongated left-tails of the peaks. In this picture the sub-condensate is a rare event whose probability tends to zero when the system size increases, as illustrated by the numerical analysis of $P_{\rm dev}$ above.

\section{Power-law weights} 
\label{sec:power-law}
\begin{figure}
    \centering
    \includegraphics[width=0.489\textwidth]{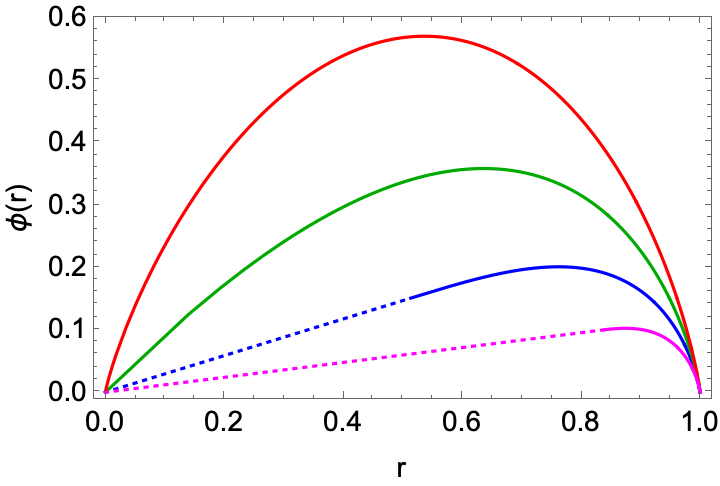}
    \includegraphics[width=0.479\textwidth]{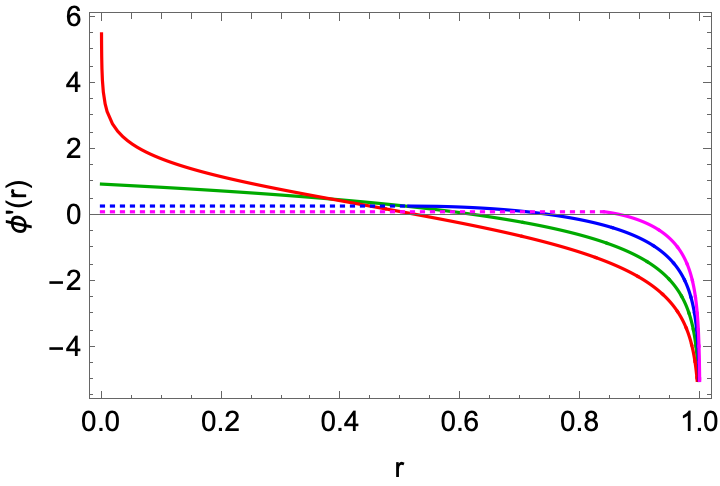}
    \caption{Left: $\phi(r)$; Right: $\phi'(r)$ for (A) $\beta=7/2$ - magenta;
    (B) $\beta=5/2$ - blue; (C) $\beta=3/2$ - green; (D) $\beta=1/2$ - red. There is 
    a phase transition for (A) and (B) at $r_{\rm c}=\zeta(\beta)/\zeta(\beta-1)$ which is
    approximately at $0.8399$ and $0.5135$, respectively.
    }
    \label{fig:phi}
\end{figure}

Consider weights of the form
\begin{equation}
\tilde{w}(s) = s^{-\beta} e^{\lambda + \sigma s} \quad \rm{for}  \quad s=1,2,\ldots 
\label{w_tls}
\end{equation}
with $\beta,\lambda,\sigma$ being arbitrary real parameters.
These weights can be reduced to power-law weights
\begin{equation}
    w(s) = s^{-\beta} \quad \rm{for}  \quad s=1,2,\ldots 
    \label{plw}
\end{equation}
by the transformation (\ref{wtransf}) $\tilde{w}(s) \rightarrow w(s) = e^{-\lambda - \sigma s} \tilde{w}(s)$. 
As follows from (\ref{OO}) the ensemble averages for the power-law weights (\ref{plw}) are the same as for the weights $\tilde{w}(s)$ (\ref{w_tls}).  We will therefore restrict ourselves to the weights (\ref{plw}), 
which are representative for the whole family (\ref{w_tls}). The generating function 
$K(\alpha)$ is
\begin{equation}
   K(\alpha) = \log {\rm Li}_\beta(e^{-\alpha}) 
   \label{kalpha}
\end{equation}
where ${\rm Li}_\beta(z)$ is the polylogarithm \cite{as}:
\begin{equation}
    {\rm Li}_\beta(z) = \sum_{k=1}^\infty \frac{z^k}{k^\beta} .
\end{equation}
The asymptotic behaviour of $K(\alpha)$ for $\alpha\rightarrow 0^+$ can be deduced from
the asymptotic behavior of the polylogarithm which is discussed in
Appendix \ref{app:polylog}. It depends on $\beta$.
Regarding the critical behaviour of the model 
we can distinguish four cases: 
\begin{itemize}
    \item[(A)]  $K''(0) < \infty$; $\beta\in (3,+\infty)$;  
    \item[(B)]  $K''(0) = \infty$ but $K'(0) < \infty$; $\beta\in (2,3]$; 
    \item[(C)]  $K'(0) = \infty$ but $K(0) < \infty$; $\beta\in (1,2]$; 
    \item[(D)]  $K(0) = \infty$; $\beta \in (-\infty,1]$;
\end{itemize}
{\label{color} which lead to different types  
of the free energy density behaviour.} 
The free energy density $\phi(r)$ (\ref{phi}) can be calculated using the saddle point equations (\ref{sp_phi}) and (\ref{alphastar}) which lead to a parametric representation of $\phi(r)$
\begin{equation}
    r = -\frac{1}{K'(\alpha)}, \ 
    \phi(r) = \alpha - \frac{K(\alpha)}{K'(\alpha)}
    \label{pr}
\end{equation}
with the parameter $\alpha$ which varies in the range $(0,\infty)$.
These equations hold for $r > r_{\rm c}$.
The critical value is 
\begin{equation}
r_{\rm c} = - \frac{1}{K'(0)} =\frac{\zeta(\beta)}{\zeta(\beta-1)}
\end{equation}
for (A) and (B) and $r_{\rm c}=0$ for (C) and (D).
For $0<r\le r_{\rm c}$ the free energy grows linearly with $r$
\begin{equation}
    \phi(r)= r K(0) = r \log \zeta(\beta).
    \label{linearpart}
\end{equation}
The behaviour of $\phi(r)$ is illustrated
in Fig. \ref{fig:phi} for $\beta=7/2,5/2,3/2,1/2$ which are representative for the four cases. The curves are obtained by the parametric equations (\ref{pr}).
The linear part of the solution (\ref{linearpart}), which corresponds to 
the condensed phase, is shown in dashed line.
The derivative $\phi'(r)$ is shown in the right chart in
Fig. \ref{fig:phi}. It is calculated from Eq. (\ref{sp_phi}) which gives $\phi'(r) = K(\alpha(r))$. In combination with (\ref{alphastar})
this leads to the following parametric equations for the derivative
\begin{equation}
    r= -\frac{1}{K'(\alpha)}, \    \phi'(r) = K(\alpha)
    \label{phi_prim_par}
\end{equation}
for $r > r_{\rm c}$, with the parameter $\alpha$ which varies in the range 
$(0,\infty)$.
For $0< r \le r_{\rm c}$ the derivative is constant
\begin{equation}
    \phi'(r)=K(0)
\end{equation}
as follows from (\ref{pr}) and (\ref{linearpart}). For (A)
and (B) the derivative $\phi'(r)$ is constant for $0\le r \le r_{\rm c}$.
For $r\rightarrow 1$ the derivative $\phi'(r)$ tends to $-\infty$ . 
For $r\rightarrow 0$ it tends to $K(0)$ which is finite for (A-C) and 
infinite for (D). The main difference between (A) and (B) is that the second derivative $\phi''(r)$ is continuous at $r_{\rm c}$ for (B) and discontinuous for (A). 
More precisely (see Appendix \ref{app:s_fenergy}), when the critical point is approached from the fluid phase side, $r \rightarrow r_{\rm c}^+$, the second derivative $\phi''(r)$ behaves at the critical point $r_{\rm c}$ as follows:
\begin{equation}
\phi''(r) = \left\{ \begin{array}{lcl} 
-c_1 (r- r_{\rm c})^{x} + \ldots
& {\rm for} & \beta \in (2,3) \\
+c_2 \log (r-r_{\rm c}) + \ldots & {\rm for} & \beta =3 \\
-c_3 + \ldots & {\rm for} & \beta \in (3,\infty)
\end{array} \right.
\end{equation}
where 
\begin{equation}
x=\frac{3-\beta}{\beta-2}     
\end{equation}
is a positive exponent and $c_1,c_2,c_3$ are positive
constants that depend on $\beta$. Dots indicate sub-leading terms.
On the other hand, the second derivative is zero $\phi''(r)=0$ 
in the condensed phase, that is for $r<r_{\rm c}$. This means that the phase transition is second order for 
$\beta\in (3,+\infty)$, with  a finite discontinuity of $\phi''(r)$ 
at the critical value $r=r_{\rm c}$. For $\beta=3$, the phase transition is also 
second order but $\phi''(r)$ has a logarithmic discontinuity 
$\log (r-r_{\rm c})$, as $r$ approaches $r_{\rm c}$ from above. 
For $\beta\in (2,3)$, the phase transition is of third or higher order. 
The order of the transition increases when $\beta$ approaches $2$ and the transition
eventually disappears for $\beta=2$. There is no phase transition for $\beta\le 2$.

\section{Grand-canonical ensemble}\label{sec:GC}

So far we have considered a system of $S$ particles in $N$ boxes and analysed
its behaviour in the thermodynamic limit (\ref{dlimit}) as a function of the 
limiting particle density $\rho = 1/r = S/N$. Now we will consider a system
with a variable number of boxes $N$, which is controlled by 
``chemical potential'' $\mu$, which is equal 
to the energy that can be absorbed or released in the system due to a change of 
the number of boxes by one. The corresponding partition function is
\cite{bbj2,g3}
\begin{equation}
    Z_{S,\mu} = \sum_{N=1}^S e^{-\mu N} Z_{S,N}.
    \label{ZSmu}
\end{equation}
The system described by the partition function (\ref{ZSmu})
will be called grand canonical ensemble. 

The grand-canonical averages are defined as
\begin{equation}
    \langle O \rangle_{S,\mu} = \frac{1}{Z_{S,\mu}} \sum_{N=1}^S e^{-\mu N} Z_{S,N} \langle O \rangle_{S,N} .
    \label{gcO}
\end{equation}
A general comment on notation: we will distinguish between canonical averages 
and grand-canonical ones by the subscripts of the brackets 
which will be $S,N$ in the first case, and $S,\mu$ in the second.

As before, let us discuss the particle distribution. 
The grand-canonical average is
\begin{equation}
    \langle \pi(q) \rangle_{S,\mu} = \frac{w(q)}{Z_{S,\mu}} \sum_{N=1}^{S-q+1} e^{-\mu N} Z_{S-q,N-1} ,
    \label{piq1}
\end{equation}
as directly follows from the definition (\ref{gcO}). The contribution to the sum from $N=1$ has to be treated
carefully, because if there is only one box it must contain all particles. Therefore for $N=1$ the expression 
$Z_{S-q,N-1}$ on the right hand side (\ref{piq1}) should be 
interpreted as $Z_{S-q,0} = \delta_{S-q}$ (see Eq. (\ref{ZS0})).
Taking this into account we can rewrite the last equation as
\begin{equation}
    \langle \pi(q) \rangle_{S,\mu} = e^{-\mu} w(q) \frac{Z_{S-q,\mu}}{Z_{S,\mu}} + e^{-\mu} w(S)
    \frac{\delta_{S-q}}{Z_{S,\mu}}
    \label{piq2}
\end{equation}
for $q=1,2,\ldots, S$. We are now interested in the behaviour of grand-canonical averages in
the thermodynamic limit, $S\rightarrow \infty$. To describe properties of the system 
in this limit we define the following thermodynamic potential
\begin{equation}
 \psi(\mu) = \lim_{S\rightarrow \infty} \frac{1}{S} 
 \log Z_{S,\mu} .
\label{psi}
\end{equation}
Again, with a slight misuse of terminology, $\psi(\mu)$ can be called grand potential or the Landau free energy density,
in analogy to the free energy density 
$\phi(r)$ (\ref{phi}) which was defined for the
canonical ensemble. The two thermodynamic potentials are related to each
other by the Legendre-Fenchel transform
\begin{equation}
    \psi(\mu) = 
    \sup_{r \in [0,1]} \left(-\mu r + \phi(r)\right) .
    \label{psi_phi}
\end{equation}
Indeed, approximating the grand-canonical partition function (\ref{ZSmu}) by an integral and extracting its leading exponential behaviour (\ref{spZ})
\begin{equation}
    Z_{S,\mu} \propto e^{S\psi(\mu)}
    \propto \int_0^1 dr e^{S\left(-\mu r + \phi(r)\right)}
    \label{saddle_point}
\end{equation}
leads to (\ref{psi_phi}). The inverse transform is
\begin{equation}
    \phi(r) = \sup_{\mu \in R} \left(\mu r + \psi(\mu)\right) .
\end{equation}
In most cases it reduces to 
\begin{equation}
    \phi(r) = \mu(r) r + \psi(\mu(r))
    \label{lf_phi}
\end{equation}
with $\mu(r)$ being a solution of
\begin{equation}
    r = -\psi'(\mu) .
    \label{rpsi}
\end{equation}
Comparing  (\ref{sp_phi}) and (\ref{lf_phi}) we find the following consistency equations
\begin{align}
   \mu = K(\alpha) \label{e1} \\
   \alpha = \psi(\mu) 
   \label{e2}
\end{align}
from which we can deduce that the grand-canonical
thermodynamic potential $\psi(\mu)$ (\ref{psi}) is the inverse function of the cumulant generating function (\ref{K}) $\psi(\mu) = K^{-1}(\mu)$, or equivalently that 
\begin{equation}
\psi(K(\alpha)) = \alpha , \quad \ K(\psi(\mu)) = \mu.
\label{psiK}
\end{equation}
It follows that $K'(\psi(\mu)) = 1/\psi'(\mu)$, which shows that
also Eqs. (\ref{alphastar}) and (\ref{rpsi})
are consistent. Using this exact relation we can calculate 
the average $r=r(\mu)$ in the thermodynamic limit $S\rightarrow \infty$
in the grand-canonical ensemble 
\begin{equation}
  r(\mu) = \lim_{S\rightarrow \infty}
  \frac{ \left\langle  N   \right\rangle_{S,\mu}  }{S} = 
  - \psi'(\mu) = \frac{-1}{K'\left(K^{-1}(\mu)\right)}  .
\label{rho_mu}
\end{equation}
This result holds for $\mu \le \mu_{\rm c}$, where $\mu_{\rm c}$ is
the critical value of the chemical potential given by
\begin{equation}
\mu_{\rm c} = K(\alpha_{\rm c}).
\label{mu_cr}
\end{equation}
where $\alpha_{\rm c}$ is given by (\ref{alpha_min}). 
For power-law weights (\ref{plw}) $\alpha_{\rm c}=0$. 
When the chemical potential exceeds the critical value, $\mu>\mu_{\rm c}$, 
the saddle point equation breaks down and the average $r=r(\mu)$ (\ref{rho_mu}) drops to zero. 
In this case $N$ grows sub-linearly with $S$ for $S\rightarrow \infty$.

The box-occupation probability in the fluid phase in 
the grand-canonical ensemble can be calculated in the thermodynamic limit as follows. For large $S$  ($S \gg 1$) the partition function $Z_{S,\mu}$ (\ref{saddle_point}) can be approximated by $e^{S\psi(\mu)}$ (\ref{saddle_point}). 
Substituting this into (\ref{piq2}) we get 
\begin{equation}
 \langle \pi(q) \rangle_{S,\mu} = w(q) e^{-q \psi(\mu) - \mu} .
 \label{exp2_sup}
\end{equation}
Using the consistency relations (\ref{e1}) and (\ref{e2}) we
see that the canonical  and grand-canonical averages
of the box-occupation probability (\ref{exp_sup}) and (\ref{exp2_sup}) 
are identical
\begin{equation}
\langle \pi(q) \rangle_{S,\mu}= \langle \pi(q) \rangle_{S,N}
\ {\rm for} \ S \rightarrow \infty 
\end{equation}
when the particle density in the canonical ensemble
is related to the chemical potential in the grand-canonical ensemble as follows $N/S =  -K'(K^{-1}(\mu))$ (\ref{rho_mu}).
The equivalence holds in the saddle point regime (fluid phase), that is for $\mu< \mu_{\rm c}$ (\ref{mu_cr}). 
It breaks down at $\mu = \mu_{\rm c}$. For $\mu>\mu_{\rm c}$ 
the average number of boxes
$\langle N \rangle_{S,\mu}$ approaches a constant independent of $S$
for $S\rightarrow \infty$. Systems with a constant number of boxes and the number of particles tending to infinity exhibit a condensation of almost all particles in a single box \cite{g3}.

\begin{figure}
    \centering
    \includegraphics[width=0.479\textwidth]{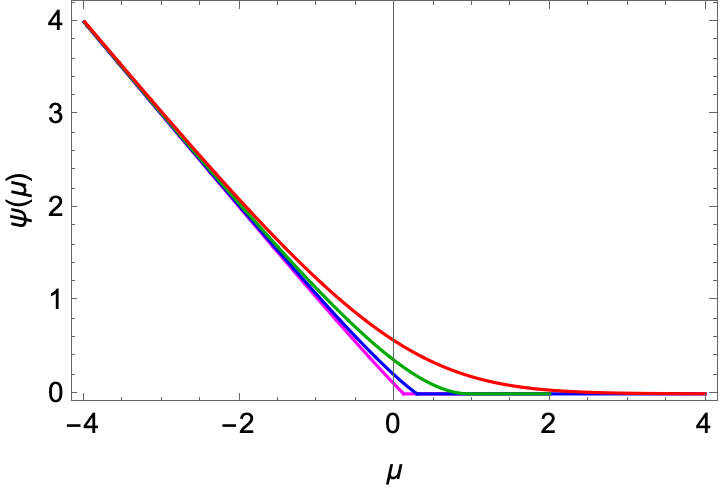}
    \includegraphics[width=0.489\textwidth]{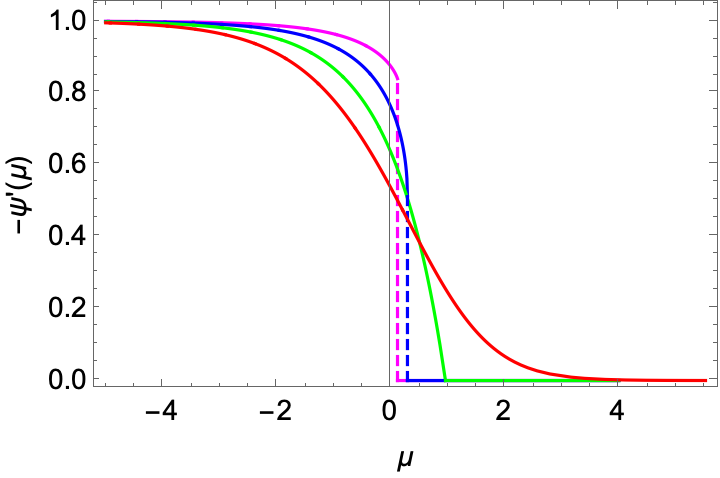}
    \caption{Left: $\psi(\mu)$; Right: $\psi'(\mu)$ for (A) $\beta=7/2$ - magenta;
    (B) $\beta=5/2$ - blue; (C) $\beta=3/2$ - green; (D) $\beta=1/2$ - red.
    The critical values $\mu_{\rm c}=\log \zeta(\beta)$ are approximately
    $0.1193$, $0.2938$ and $0.9603$ for (A), (B) and (C),
    respectively. Dashed straight lines indicate the discontinuities size
    $r_{\rm c} =\zeta(\beta)/\zeta(\beta-1)$ of $\psi'(\mu)$ at $\mu_{\rm c}$ which are $0.8399$ and $0.5135$ for (A) and (B), respectively.}
    \label{fig:psi}
\end{figure}

To illustrate the different types of possible behaviour 
of the Landau free energy density $\psi(\mu)$, let us draw curves representing 
$\psi(\mu)$ and $\psi'(\mu)$ for power-law weights (\ref{plw}) for
different $\beta$. As for the free energy density
(see Fig. \ref{fig:phi}) we will show in each plot four curves  
for (A) $\beta=7/2$, (B) $\beta=5/2$,  (C) $\beta=3/2$, and (D) $\beta=1/2$.
The curves will be generated as parametric plots. To plot $\psi(\mu)$ we
use the fact that $\psi$ is the inverse function of $K$ (\ref{psiK})
which is equivalent to the following parametric equation
\begin{equation}
    \mu = K(\alpha) \ , \quad \psi(\mu) = \alpha
\end{equation}
with the parameter $\alpha>\alpha_{\rm c}$, which for power-law weights 
(\ref{plw}) $\alpha_{\rm c}=0$. For $\psi'(\mu)$ we have 
the following parametric equations:
\begin{equation}
    \mu = K(\alpha) \ , \
    \psi'(\mu) = \frac{1}{K'(\alpha)}
    \label{psi_prim_par}
\end{equation}
which are a direct consequence of $\psi$ being the inverse function of $K$.
The results are shown in Fig. \ref{fig:psi}.
For power-law weights (\ref{plw}) the critical value of the chemical potential 
is 
\begin{equation}
   \mu_{\rm c}=K(0)=\log \zeta(\beta).
   \label{mu_cr_plw}
\end{equation}
Note that the parametric equations for $\psi'(\mu)$ (\ref{psi_prim_par})
are almost identical to those for $\phi'(r)$ (\ref{phi_prim_par}). There are two
differences. The first is that there is a minus sign in front 
of $1/K'(\alpha)$ in the equations for $\psi'(r)$ which is absent in the equations for $\psi'(\mu)$. The second is that the right-hand sides of these
equations are swapped which means that the ordinate and of abscissa 
switch roles. This is not surprising because $\psi(r)$ and $\phi(\mu)$ 
are related to each other by the Legendre-Fenchel transformation. 
In other words the drawings of $\psi'(\mu)$ and $\phi'(r)$ (compare
Fig.\ref{fig:psi} and Fig.\ref{fig:phi}) can be obtained from each 
another by swapping the vertical and horizontal axis, and changing the direction
of the vertical axis.

So far we have discussed the case of power-law weights (\ref{plw}). The question is
what the corresponding plots look like for the family of weights (\ref{w_tls}). Again this
can answered using the transformation $w(s) \rightarrow e^{\lambda + \sigma s} w(s)$ (\ref{wtransf}). Under this transformation the
free energy density (\ref{phi}) transforms as 
\begin{equation}
    \phi(r) \rightarrow \phi(r) + \sigma + r \lambda
\end{equation}
and grand potential transforms as
\begin{equation}
    \psi(\mu) \rightarrow \psi(\mu-r) + \sigma 
\end{equation}
as follows from (\ref{psi_phi}). The derivatives transform as
\begin{equation}
    \phi'(r) \rightarrow \phi'(r) + \lambda
\end{equation}
and 
\begin{equation}
    \psi'(\mu) \rightarrow \psi'(\mu-\lambda),
\end{equation}
respectively. This means that the curve $\phi'(r)$ (see Fig. \ref{fig:phi})
moves up by $\lambda$ and the curve $\psi'(\mu)$ (see Fig. \ref{fig:psi})
moves right by $\lambda$ when power-law weights (\ref{plw}) are changed to (\ref{w_tls}). 
The critical value $r_{\rm c}$ (\ref{rcr}) does not change. 
The critical value  $\mu_{\rm c}$ (\ref{mu_cr_plw}) 
is shifted by $\lambda$ from
$\log \zeta(\beta)$ to
\begin{equation}
    \mu_{\rm c} = \log \zeta(\beta) + \lambda .
    \label{mu_shifted}
\end{equation}
The parameter $\sigma$ has no effect on the critical values $r_{\rm c}$ and $\mu_{\rm c}$.

Returning to purely power-law weights let us now concentrate on the phase
transition, i.e. on the behaviour of the grand potential $\psi(\mu)$
near the critical point $\mu_{\rm c}$ (\ref{mu_cr_plw}). The detailed
analysis is presented in Appendix \ref{app:s_grand_pot}. 
For $\mu \rightarrow \mu_{\rm c}^-$ the grand potential behaves as
\begin{equation}
    \psi'(\mu) = \left\{ \begin{array}{lcl} 
-C (\mu_{\rm c} - \mu)^y + \ldots 
& {\rm for} & \beta \in (1,2) \\
 \zeta(2)/\log(\mu_{\rm c}-\mu) + \ldots
 & {\rm for} & \beta =2 \\
-r_{\rm c} + \ldots & {\rm for} & \beta \in (2,+\infty) 
\end{array} \right.
\end{equation}
where 
\begin{equation}
y = \frac{2-\beta}{\beta-1} 
\end{equation}
is a positive exponent and $C$ is a positive constant depending on $\beta$. Dots indicate sub-leading terms.
On the other hand, $\psi'(\mu)=0$ for $\mu>\mu_{\rm c}$.
This means that the phase transition is  first order for $\beta\in (2,+\infty)$. 
For $\beta\in (1,2]$ the order of the phase transition varies 
from second to infinite. More precisely, the interval $\beta\in (1,2]$
can be divided into sub-intervals $\beta \in (1+1/n,1 + 1/(n-1)]$, $n=2,3,\ldots$, in which the $n$-th derivative is discontinuous. The transition disappears for $\beta=1$.

\section{Other  ensembles \label{sec:oe}}

One can consider a system with $N$ boxes 
and a variable number of particles.
The following partition function 
\begin{equation}
    Z_{\alpha,N} = \sum_{S=N}^\infty e^{-\alpha S} Z_{S,N} .  
\end{equation}
describes the possible states of such a system in equilibrium with a common reservoir that each box is in contact with.
In the thermodynamic limit the reservoir is assumed to contain infinitely many particles. The fugacity $\alpha$ plays the role of the chemical potential of a particle. It is easy to see that the partition function factorises in this case
\begin{equation}
    Z_{\alpha,N} = e^{N K(\alpha)} 
\end{equation}
so it describes $N$ independent boxes, each contributing an energy $-K(\alpha)$
to the total energy of the system. The contributions from different boxes are thus entirely independent and the partition function factorises into the product of single urn/box  partition functions \cite{gs,rs}.
The average particle density is 
$\rho(\alpha) = -K'(\alpha)$. For power-law weights (\ref{plw})
with $\beta\le 1$, the density $\rho(\alpha)$ becomes infinite for 
$\alpha \le 0$. For $\beta>1$, the density 
$\rho(\alpha)$ becomes infinite for $\alpha<0$. This infinite density means that each box absorbs infinitely many particles from the reservoir. 

In principle, also an ensemble with varying numbers of particles and boxes can be
defined by the following partition function
\begin{equation}
    Z_{\alpha,\mu} = \sum_{N=1}^\infty e^{-\mu N} Z_{\alpha,N} =  
    \sum_{N=1}^\infty e^{(-\mu + K(\alpha)) N}
    \label{ggz}
\end{equation}
which is convergent for $\mu>K(\alpha)$ and $\alpha>\alpha_{\rm c}$, 
yielding
\begin{equation}
    Z_{\alpha,\mu} = \frac{1}{e^{\mu - K(\alpha)} - 1} .
\end{equation}
This expression becomes infinite for $\mu \rightarrow K(\alpha)$, the sum (\ref{ggz}) diverges. From this
partition function one can also calculate the grand-canonical
partition function $Z_{S,\mu}$ (\ref{ZSmu}): 
\begin{equation}
    \frac{1}{e^{\mu - K(\alpha)} - 1} = 
    \sum_{S=1}^\infty e^{-\alpha S} Z_{S,\mu}.
\end{equation}
as an inverse Laplace transform. 

\section{Probabilistic and quasi-probabilistic normalisation of weights}
\label{sec:quasi}

In some problems it is convenient to normalise the weights
$w(s)$ in the partition function (\ref{ZSN}) 
 \begin{equation}
   \widehat{w}(s) = \frac{w(s)}{\sum_{q=1}^\infty w(q)} .
   \label{nw}
\end{equation}
so that the model after the normalisation can be interpreted
in a probabilistic manner.
This can be done only if the sum in the denominator 
on the right hand side in (\ref{nw}) is finite.
The normalisation can be obtained 
by the transformation (\ref{wtransf}) 
with $\sigma=0$ and $\lambda = -\log \left(\sum_{q=1}^\infty w(q)\right)$ and therefore it  has no effect on the statistical averages (\ref{OO}). 

There are however families of weights which cannot be 
normalised because the normalisation sum in (\ref{nw}) is infinite. For example, the constant weights $w(s)=1$ for $s=1,2,\ldots$ cannot be normalised.
As a way round this problem, some authors have proposed
a {quasi}-probabilistic normalisation \cite{maac} by putting an upper limit on the weights (and hence sum) 
\begin{equation}
\widehat{w}(s,S) = \frac{w(s)}{\sum_{s=1}^S w(s)} = 
 \frac{w(s)}{\Omega(S)}
 \label{omega}
\end{equation}
for $s=1,2,\ldots, S$, and $\widehat{w}(s,S)=0$ for $s>S$. 
The normalisation constant $\Omega(S)$ depends on $S$.
It is finite for any finite $S$. For example, $\Omega(S)=S$
for the trivial weights $w(s)=1$. This type of cut-off grand-canonical ensemble has also been introduced as a way to control finite-size effects for the urn model with weights that lead
to a slow convergence to the thermodynamic limit \cite{cg1}.

The model with the quasi-probabilistic normalisation
can be obtained from the original weights
by the transformation (\ref{wtransf}) with 
$\lambda=-\log \Omega(S)$. This leads to the following
relationship between the partition functions
\begin{equation}
    \widehat{Z}_{S,N} = e^{-N \Omega(S)} Z_{S,N}
\end{equation}
and 
\begin{equation}
    \widehat{Z}_{S,\mu} = Z_{S,\mu + \log \Omega(S)} .
    \label{ZhatZ}
\end{equation}
where the partition functions on the right hand side of the
equations above are defined by weights independent of $S$,
see Eqs. (\ref{ZSN}) and (\ref{ZSmu}).

We note that the model with the quasi-probabilistic normalisation
(\ref{omega}) is no longer invariant with respect to 
the transformation (\ref{wtransf}), because $\Omega(S)$
changes under this transformation. It therefore makes a difference
whether one considers purely power-law weights $q^{-\beta}/\Omega(S)$ or exponentially damped ones $q^{-\beta} e^{-\sigma q}/\Omega(S)$. 
We stress that the weights $w(s)$ in the partition functions
(\ref{ZSN}) and (\ref{ZSmu}) depend only on the occupation of the box while in the quasi-probabilistic model the weights 
$\widehat{w}(s,S)$ depend also on $S$.  
The quasi-probabilistic model therefore belongs to a different
class.

In the rest of this section we will consider
the quasi-probabilistic model with the power-law weights 
(\ref{plw}) in the grand-canonical ensemble for $\mu=0$, which corresponds to the model discussed in the main part of this paper (\ref{ZSmu}) with a running chemical potential
\begin{equation}
\mu(S) = \log \Omega(S)= \log \sum_{s=1}^S s^{-\beta}.
\label{mu_omega}
\end{equation}
This follows from (\ref{ZhatZ}). 
For $\beta>1$, the running chemical potential $\mu(S)$ approaches the critical value
$\mu_{\rm c}= \log \zeta(\beta)$ from below $\mu(S) \rightarrow \mu_{\rm c}^-$ when $S$ increases. The difference $\mu_{\rm c}-\mu(S)$ behaves as
\begin{equation}
    \mu_{\rm c} - \mu(S) = 
    \log \left(1 - \frac{\zeta(\beta,S+1)}{\zeta(\beta)}\right) .
\end{equation}
where $\zeta(\beta,S)$ is the Hurwitz zeta function \cite{as}.
The difference tends to zero as 
\begin{equation}
   \mu_{\rm c} - \mu(S) = \frac{1}{(\beta-1)\zeta(\beta)}S^{-
   (\beta-1)} + \ldots 
   \label{mS}
\end{equation}
for $S \rightarrow \infty$.  Dots indicate sub-leading terms.
We can compute the average inverse density at $\mu(S)$
using the equation (\ref{rho_mu}) for $S \gg 1$
\begin{equation}
    r(S)=\frac{\langle N \rangle_{S,\mu(S)}}{S} = -\psi'\left(\mu(S)\right) .
    \label{rN}
\end{equation}
For $\beta\in (2,\infty)$, $r(S)$ tends to the critical value  
$\zeta(\beta)/\zeta(\beta-1)$ for $S\rightarrow \infty$. For $\beta \in (1,2)$, $r(S)$ behaves asymptotically as
$r(S) \sim S^{\beta-2}$ for large $S$, as can be seen by substituting (\ref{mS}) into (\ref{psi_critical_tau_in_1,2}).
For $\beta \in (-\infty,1]$ there is no phase transition. In this case $\mu(S)$ (\ref{mu_omega}) grows asymptotically as
$\mu(S) \sim (-\beta+1) \log(S)$. Substituting this into 
(\ref{psi_prim_exponential}) we see that $r(S)$ 
behaves asymptotically as $r(S) \sim 1/S$ for large $S$.
These results translate into (\ref{rN})
\begin{equation}
\langle N \rangle_{S,\mu(S)} = 
\left\{ 
\begin{array}{lll} 
c_1 S + \ldots & \mbox{for} & \beta \in (2,\infty) \\
c_2 S^{\beta-1} + \ldots & \mbox{for} & \beta \in (1,2) \\
c_3 + \ldots  & \mbox{for} & \beta \in (-\infty,1) 
\end{array}
\right.
\end{equation}
for $S\rightarrow \infty$, with some positive coefficients $c_1$, $c_2$ and $c_3$ depending on $\beta$,  where dots again indicate sub-leading terms.
The detailed results, also for the borderline cases of
$\beta=1$ and $\beta=2$, can be found in \cite{maac}.

Let us stress that the model with weights
normalised by the pseudo-probabilistic condition (\ref{omega}) corresponds to the model discussed in the main part of 
this paper (\ref{ZSmu}) with a running chemical potential $\mu=\log \Omega(S)$ (\ref{mu_omega}).
For $\beta>1$, the effective chemical potential 
$\mu=\log \Omega(S)$ approaches the critical
value $\mu_{\rm c}=\log \zeta(\beta)$ from below as $S$
increases. For $\beta \le 1$, the
effective chemical potential tends logarithmically to infinity, 
for $S\rightarrow \infty$. In both cases the system stays in the fluid phase, as long as $S$ is finite. 
The effective particle probability distribution (\ref{exp2_sup}) 
for the system with the effective chemical potential $\mu=\log \Omega(S)$ is 
\begin{equation}
 \langle \pi(q) \rangle_{S,\log \Omega(S)} = 
 \frac{w(q)}{\Omega(S)} e^{-q \psi(\log \Omega(S))} .
\end{equation}
We thus see that the quasi-probabilistic normalisation introduces 
weak exponential damping for large $q$ into the effective particle distribution. The damping factor decays with $S$ but disappears completely only in the limit $S\rightarrow \infty$.

\section{Summary}\label{sec:summary}

After some general discussion of the urn model
we have focused in this paper on the zeta-urn model which  
is analytically solvable. 
We have used this to determine the thermodynamic potentials that control the
exponential growth of the canonical and grand-canonical partition
functions in the thermodynamic limit and elucidate the critical behaviour. 

The second derivative of the free energy density with respect to the particle density describes density fluctuations in the canonical 
ensemble. For $\beta>3$ the second derivative is discontinuous 
at the critical point, so the transition is of the second order. 
For $\beta=3$ the second derivative has a logarithmic divergence. 
For $\beta \in (2,3]$ the order of the transition changes   
at the discrete values $\beta_n=2+1/(n-1)$, $n=2,3,\ldots$,
where $n$ is the order of the transition for $\beta \in (\beta_{n+1},\beta_n]$. There is no phase transition for $\beta\le 2$. 

For the grand-canonical ensemble the situation is different. 
The first derivative of the corresponding thermodynamic 
potential is discontinuous for $\beta>2$, so the phase transition 
is of the first order in this case. 
For $\beta \in (1,2]$ the order of the transition changes  
at the discrete values $\beta'_n = 1+1/(n-1)$, where $n=2,3,\ldots$ 
is the order of the transition for  $\beta \in (\beta'_{n+1},\beta'_n]$.
There is no phase transition for $\beta\le 1$.

The thermodynamic potentials for the canonical and grand-canonical
ensembles can be derived from each other. More specifically, the function: $\phi'(r)$ on the support $r\in (r_{\rm c},1)$ is the inverse function of $\mu\rightarrow -\psi'(\mu)$ on the support $\mu \in (-\infty,\mu_{\rm c})$ which is a consequence of the Legendre-Fenchel transform which relates the two functions. Additionally, we have shown the grand-canonical potential $\psi(\mu)$ on the support 
$\mu \in (-\infty,\mu_{\rm c})$ is the inverse function of the cumulant generating function $K(\alpha)$ on the support $\alpha \in (\alpha_{\rm c},\infty)$.
The relation of other ensembles to the canonical and grand-canonical ensembles discussed in detail here has also been explored, in particular those with probabilistic and quasi-probabilistic weights which can be useful both for normalising otherwise divergent sums and for controlling finite-size effects.

It is perhaps worth emphasizing that the zeta-urn model provides a useful toy model for investigating phase transitions of any order, including discontinuous ones in the case of the grand-canonical ensemble, by the simple expedient of varying a single parameter - the exponent $\beta$ in the power law weights. The finite-size partition functions are, at least in principle, exactly calculable and can be compared with the asymptotic results of saddle point expansions to explore the finite-size effects at the transition with $\beta$ tuned to give the desired order.

In \cite{bbjr} we use some of the results presented in this paper
to study the R\'enyi entropy and diversity measures \cite{maac,wbe}
for the zeta-urn model and discuss their singular behaviour and other properties in 
the thermodynamic limit (\ref{dlimit}) in the canonical ensemble. On the other hand, the grand-canonical ensemble is employed in \cite{bbjz} to investigate the behaviour of the partition function zeros of the zeta-urn as $\beta$ is varied and the order of the transition changes.

\begin{acknowledgements}
\label{ack}
The authors would like to thank Paul Chleboun for  a helpful discussion on finite-size corrections.
\end{acknowledgements}

\appendix

\section{Derivation of the particle distribution in the thermodynamic limit} \label{app:piq}

The starting point of the calculation is Eq. (\ref{piq})
We replace the partition functions in the numerator and denominator
on the right hand side of this equation
by their asymptotic forms which follow the saddle point
approximation (\ref{spZ})
\begin{equation}
    Z_{S,N} \propto e^{S\alpha(r) + N K(\alpha(r))} 
\end{equation}
and 
\begin{equation}
    Z_{S-q,N-1} \propto  e^{(S-q)\alpha(\tilde{r}) + (N-1) K(\alpha(\tilde{r}))} 
    \label{Z2}
\end{equation}
where $r=N/S$ and $\tilde{r}=(N-1)/(S-q)$. 
In the limit (\ref{dlimit}) $\tilde{r}$ can be expanded in $1/S$:
\begin{equation}
    \tilde{r} = r + \frac{q r-1}{S} + o\left(\frac{1}{S}\right).
\end{equation}
It follows that 
\begin{equation}
    \alpha(\tilde{r}) = \alpha(r) + \alpha'(r) \frac{qr-1}{S} + 
    o\left(\frac{1}{S}\right).
\end{equation}
and 
\begin{equation}
    K(\alpha(\tilde{r})) = K(\alpha(r)) + K'(\alpha(r))\alpha'(r) 
    \frac{qr-1}{S} + o\left(\frac{1}{S}\right).
\end{equation}
Using the saddle point equation \eqref{alphastar} we can replace
$K'(\alpha(r))$ in the last equation with $-1/r$:
\begin{equation}
    K(\alpha(\tilde{r})) = K(\alpha(r)) - \alpha'(r) 
    \frac{qr-1}{r S} + o\left(\frac{1}{S}\right).
\end{equation}
Inserting these expansions into (\ref{Z2}) we get
\begin{equation}
    Z_{S-q,N-1} \propto \exp\left\{(S-q) \alpha(r) + (N-1) K(\alpha(r))
    + \alpha'(r) (qr-1)^2/(rS) + o\left(S^{-1}\right)  \right\} .
\end{equation}
and hence
\begin{equation}
    \frac{Z_{S-q,N-1}}{Z_{S,N}} \propto \exp\left\{ -q \alpha(r) - K(\alpha(r)) + \alpha'(r) (qr-1)^2/(rS)  
    + o\left(s^{-1}\right)\right\} ,
\end{equation}
and finally (\ref{piq})
\begin{equation}
  \langle \pi(q) \rangle_{S,N} \propto w(q) 
  \exp\left\{ -q \alpha(r) - K(\alpha(r))
    + O\left(S^{-1}\right)\right\} .
\end{equation}

\section{Asymptotics of the polylogarithm} \label{app:polylog}

For non-integer $\beta$, the polylogarithm of $e^{-\alpha}$ 
has the following series expansion at $\alpha_{\rm c}=0$ 
\begin{equation}
   {\rm Li}_\beta(e^{-\alpha}) = 
   \Gamma(1-\beta) \alpha^{\beta-1} + \sum_{k=0}^\infty
   \frac{(-1)^k \zeta(\beta-k)}{k!} \alpha^k .
   \label{exp-non-int}
\end{equation}
For integer $\beta$, the singular term contains a logarithmic singularity
\begin{widetext}
\begin{equation}
   {\rm Li}_\beta(e^{-\alpha}) = 
   \frac{(-1)^{\beta-1}}{(\beta-1)!}\left(H_{\beta-1}-\log(\alpha)\right) \alpha^{\beta-1} +
    \sum_{k=0,k\ne \beta-1}^\infty
   \frac{(-1)^k \zeta(\beta-k)}{k!} \alpha^k 
   \label{exp-int}
\end{equation}
\end{widetext}
where $H_n=1+1/2+\ldots+1/n$ is the $n$-th harmonic number. 

\section{Singularities of the (canonical) free energy density at the critical point} 
\label{app:s_fenergy}

In this appendix we will discuss singularities of the free energy 
density $\phi(r)$ at the critical point $r = r_{\rm c}$,
This is the point where the solid line changes to the dashed line for curves 
(A) and (B) in Fig. \ref{fig:phi}.
The analysis will be performed on the basis of parametric equations (\ref{pr}), 
which allow us to extract the singularity of the free energy
density. The main components of these equations are 
the function $K(\alpha)$ and its derivative $K'(\alpha)$, so to prepare the ground let's 
analyze the behavior of these functions for $\alpha$ close to
$\alpha_{\rm c}=0$, which correspond to $r$ close to $r_{\rm c}$. 

For the cases (D) and (C), that is for $\beta \in (-\infty,2]$ there is no phase
transition as $r_{\rm c}=0$. So we will concentrate on the range $\beta\in (2,\infty)$. For non-integer we can write
\begin{equation}
   K(\alpha) =  \mu_{\rm c} - 
   \sum_{k=1}^{\lfloor\beta-1\rfloor} a_k \alpha^k -
   A \alpha^{\beta-1} + o(\alpha^{\beta-1})
\label{K_tau>2}
\end{equation}
and
\begin{equation}
   -\frac{1}{K'(\alpha)} = r_{\rm c} + 
   \sum_{k=1}^{\lfloor\beta-2\rfloor} b_k \alpha^k +
   B \alpha^{\beta-2} + o(\alpha^{\beta-2}) .
   \label{1_Kprim_tau>2}
\end{equation}
where the symbol $\lfloor x \rfloor$ is the largest integer less or equal $x$. 
The constants $\mu_{\rm c}$ and $r_{\rm c}$ on the right-hand side of these equations stand for
$\mu_{\rm c}=K(0)$ and $r_{\rm c}= -1/K'(0)$, respectively. 
The coefficients $a_k$'s and $b_k$'s as well as $A$ and $B$ can be directly 
deduced from the expansion of $K(\alpha)$ (\ref{kalpha}). They depend on $\beta$,
but the form of this dependence is inessential for the further analysis. The signs of 
the coefficients are chosen for convenience. 
Only terms up to $\alpha^{\beta-1}$ in the first equation, 
and up to $\alpha^{\beta-2}$ in the second are displayed. 
All others are included in the little-o symbols for 
$\alpha \rightarrow 0$.

We will separately consider the ranges $\beta \in (2,3)$ and $\beta\in (3,+\infty)$,
and the borderline case $\beta=3$. We begin with $\beta \in (2,3)$.
The parametric equations (\ref{phi_prim_par}) for the derivative $\phi'(r)$
become
\begin{equation}
    \phi'(r) =  \mu_{\rm c} - a_1 \alpha + o(\alpha)
\label{phi_prim_exp}
\end{equation}
with
\begin{equation}
    r = r_{\rm c} + B \alpha^{\beta-2} + o\left( \alpha^{\beta-2}\right) .
\label{tau_minus_2}
\end{equation}
for $\alpha$ close to $\alpha_{\rm c}=0$,
as follows from (\ref{K_tau>2}) and (\ref{1_Kprim_tau>2}).
From the second equation we find that $\alpha = ((r-r_{\rm c})/B)^{1/(\beta-2)} + \ldots$
 when $r$ tends to $r_{\rm c}$ from above. 
Plugging this into the first equation, we get
\begin{equation}
    \phi'(r) - \mu_{\rm c} = - C (r-r_{\rm c})^{
    1/(\beta-2)} + o\left(r-r_{\rm c})^{
    1/(\beta-2)}\right)
\end{equation}
where $C=a_1/B^{1/(\beta-2)}$. The second derivative behaves as 
\begin{equation}
    \phi''(r) \sim (r-r_{\rm c})^{(3-\beta)/(\beta-2)} 
\end{equation}
for $r \rightarrow r_{\rm c}^+$. The 
power $(3-\beta)/(\beta-2)$ is positive for $\beta \in (2.3)$, 
so this means that $\phi''\left(r_{\rm c}^+\right) = 0$.
Also for $r\rightarrow r_{\rm c}^-$ the second derivative is equal to zero, $\phi''\left(r_{\rm c}^-\right) = 0$, hence the second derivative is continuous for $\beta \in (2,3)$. In the same way it can be checked
that the third derivative is continuous for $\beta \in (2,5/2)$ and it is discontinuous 
for $\beta\in [5/2,3)$. The transition is thus third order for $\beta\in [5/2,3)$.
More generally, the $n$-th derivative $\phi^{(n)}(r)$ is discontinuous at $r_{\rm c}$
for $\beta \in [2+1/(n-1),3)$. Therefore the transition is of $n$-th order for 
$\beta\in[2+1/(n-1),2 + 1/(n-2))$ for $n=3,4,\ldots$. 
The order of the transition increases to infinity when $\beta$ approaches two. 
Eventually at $\beta=2$, the transition disappears completely.

Now consider the range $\beta\in (3,+\infty)$. We first focus on  
$\beta\in (3,4)$. In this case Eq. (\ref{tau_minus_2}) takes the form
\begin{equation}
    r = r_{\rm c} + b_1 \alpha + B \alpha^{\beta-2}+ o(\alpha^{\beta-2}) 
\end{equation}
For $r$ close to $r_{\rm c}$ this gives 
\begin{equation}
    \alpha = \frac{r-r_{\rm c}}{b_1} - \frac{B}{b_1}
    \left(\frac{r-r_{\rm c}}{b_1}\right)^{\beta-2}+ o\left((r-r_{\rm c})^{\beta-2}\right) .
\end{equation}
Substituting this into (\ref{phi_prim_exp}) we get
\begin{equation}
\begin{split}
    \phi'(r) - \mu_{\rm c} &=
     -\frac{a_1}{b_1}(r-r_{\rm c}) + \frac{a_1B}{b_1}
    \left(\frac{r-r_{\rm c}}{b_1}\right)^{\beta-2}+\\
    &\phantom{=} o\left((r-r_{\rm c})^{\beta-2}\right) 
\end{split}    
    \label{3rd}
\end{equation}
We see that the second derivative is discontinuous at $r_{\rm c}$,
because 
\begin{equation}
    \lim_{r\rightarrow r_{\rm c}^+} \phi''(r) = -\frac{a_1}{b_1}
\end{equation}
and
\begin{equation}
\lim_{r\rightarrow r_{\rm c}^-} \phi''(r)=0
\end{equation}
and the transition is therefore second order.
We note in passing that the third derivative has an infinite discontinuity for $\beta\in (3,4)$,
coming from the term $(r-r_{\rm c})^{\beta-2}$ in (\ref{3rd}).
The calculations can be repeated for
$\beta\in (n,n+1]$, for $n=3,4,\ldots$, to see that in all these intervals
the second derivative has a finite discontinuity and that the derivative $\phi^{(n)}(r)$ 
has an infinite discontinuity at $r_{\rm c}$.

Now consider the borderline case $\beta=3$. The parametric 
equations (\ref{phi_prim_par}) can be written as
\begin{equation}
    \phi'(r) =  
    \mu_{\rm c} - a_1\alpha + o(\alpha)
\end{equation}
and 
\begin{equation}
    r = r_{\rm c} + b_1\alpha + B\alpha\log \alpha + o(\alpha \log \alpha) ,
\end{equation}
as follows from (\ref{exp-int}). 
Inverting the second equation for $\alpha$ 
we find
\begin{equation}
\begin{split}
    \alpha &= c_1(r - r_{\rm c}) + C (r-r_{\rm c}) 
    \log (r-r_{\rm c}) +\\
    &\phantom{==} o\left((r-r_{\rm c}) 
    \log (r-r_{\rm c})\right)
\end{split}    
\end{equation}
where $c_1 = 1/b_1 + B/b_1^2 \log b_1$ and $C=-B/b_1^2$.
Substituting this into the first equation we get
\begin{equation}
\begin{split}
    \phi'(r) - \mu_{\rm c} &= - a_1 c_1 (r-r_{\rm c}) 
    - a_1 C (r-r_{\rm c}) \log (r-r_{\rm c}) + \\
&\phantom{==}    o\left((r-r_{\rm c}) \log (r-r_{\rm c})\right) .
\end{split}    
\end{equation}
Taking the derivative of both sides we find that
the second derivative has a logarithmic singularity for
$r\rightarrow r_{\rm c}^+$
\begin{equation}
\begin{split}
    \phi''(r) &= -a_1(c_1 + C) -
    a_1 C \log (r-r_{\rm c}) + \\
    &\phantom{==}o\left(\log (r-r_{\rm c})\right) . 
\end{split}    
\end{equation}

\section{Singularities of the grand potential at the critical point} \label{app:s_grand_pot}

In this appendix will determine singularities of grand potential
$\psi(\mu)$ for the power-law weights (\ref{plw}) 
at the critical chemical potential $\mu_{\rm c}=\log \zeta(\beta)$ for $\beta>1$. There is no phase transition for $\beta\le 1$.

The initial point of the analysis is the parametric representation (\ref{psi_prim_par}).
The critical behaviour is reproduced in the limit $\alpha \rightarrow 0^+$, which corresponds to $\mu_{\rm c} - \mu \rightarrow 0^+$. 
We separately consider three cases: $\beta \in (2,\infty)$, $\beta=2$ and $\beta \in (1,2)$.

For $\beta\in (2,+\infty)$ we can use Eqs. (\ref{K_tau>2}) and (\ref{1_Kprim_tau>2})
to cast the parametric equations (\ref{psi_prim_par}) into the form
\begin{equation}
\psi'(\mu) = 
- r_{\rm c} - 
   \sum_{k=1}^{\lfloor\beta-2\rfloor} b_k \alpha^k -
   B \alpha^{\beta-2} + o(\alpha^{\beta-2})
\end{equation}
and 
\begin{equation}
\mu = \mu_{\rm c} - a_1 \alpha + o(\alpha) .
\end{equation}  
From the second equation we can calculate
$\alpha = (\mu_{\rm c} - \mu)/a_1 +o(\mu_{\rm c}-\mu)$ for 
$\mu \rightarrow \mu_{\rm c}^-$. 
Substituting this into the first equation we see that
\begin{equation}
    \lim_{\mu\rightarrow \mu_{\rm c}^-} \psi'(\mu) = -r_{\rm c} ,
\end{equation}
which is a negative number. On the other hand
\begin{equation}
    \lim_{\mu\rightarrow \mu_{\rm c}^+} \psi'(\mu) = 0 ,
\end{equation}
so the first derivative has a discontinuity at $\mu=\mu_{\rm c}$, so the phase transition 
is of the first order. In addition, the first derivative has a power-law singularity 
$(\mu_{\rm c}-\mu)^{\beta-2}$, (or $(\mu_{\rm c}-\mu)^{\beta-2} \log (\mu_{\rm c}-\mu)$
for integer $\beta$), for $\mu \rightarrow \mu_{\rm c}^-$. 
This singularity makes the $n$-th and higher derivatives 
infinite for $\mu \rightarrow \mu_{\rm c}^-$,
where $n=\lfloor \beta \rfloor$. 

For $\beta=2$ there is no discontinuity of the first derivative.
In this case the first derivative behaves as 
\begin{equation}
\psi'(\mu) \sim \frac{\zeta(2)}{\log \left( \zeta(2) (\mu_{\rm c} - \mu)\right)}
\end{equation}
as $\mu$ approaches $\mu_{\rm c}$, so it vanishes in the limit $\mu\rightarrow \mu_{\rm c}^-$.
Taking the derivative of both sides we see
that the second derivative diverges as 
$\psi''(\mu) \sim (\mu_{\rm c}-\mu)^{-1} 
\left(\log(\mu_{\rm c}-\mu)\right)^{-2}$, as $\mu$ tends to $\mu_{\rm c}$,
so the transition is of the second order for $\beta = 2$.

For $\beta \in (1,2)$ the expressions (\ref{K_tau>2}) and (\ref{1_Kprim_tau>2}) reduce to
\begin{equation}
   K(\alpha) =  \mu_{\rm c} - A \alpha^{\beta-1} + o(\alpha^{\beta-1}) 
\label{K_tau_in_1,2}
\end{equation}
and
\begin{equation}
   -\frac{1}{K'(\alpha)} = 
   B \alpha^{2-\beta} + o(\alpha^{\beta-2}) .
    \label{1_Kprim_tau_in_1,2}
\end{equation}
Therefore, for $\beta \in (1,2)$ the parametric equations (\ref{psi_prim_par}) take the form
\begin{equation}
    \psi'(\mu) = -B \alpha^{2-\beta} + o(\alpha^{2-\beta})
\end{equation}
and 
\begin{equation}
    \mu_{\rm c} -\mu = A \alpha^{\beta-1} + o(\alpha^{\beta-1})
\end{equation}
for $\alpha \rightarrow 0^+$.
Extracting $\alpha$ from the second equation and substituting it into in to the first one leads to \cite{bbj2}
\begin{equation}
    \psi'(\mu) = -C (\mu_{\rm c}-\mu)^{\frac{2-\beta}{\beta-1}} + 
    o\left((\mu_{\rm c}-\mu)^{\frac{2-\beta}{\beta-1}}\right)
    \label{psi_critical_tau_in_1,2}
\end{equation}
where $C=B/A^{1/(\beta-1)}$.  
The transition is of the second order for  $\beta\in (3/2,2]$,
of the third order for $\beta \in (3/4,3/2]$, 
and of the $n$-th order for $\beta \in (1+1/n,1+1/(n-1)]$. 
The transition softens when $\beta$ approaches $1$, and it eventually disappears for $\beta=1$. 

There is no phase transition for $\beta\in (-\infty,1]$.
This can be seen directly from the expansions of $K(\alpha)$
and $K'(\alpha)$, which in this range $\beta$ instead of (\ref{K_tau>2}) and (\ref{1_Kprim_tau>2}) take the form
\begin{equation}
   K(\alpha) =  \log\left(\Gamma(1-\beta)\right) + (\beta-1) \log \alpha + o(\log \alpha) 
   \label{K_tau<1}
\end{equation}
and 
\begin{equation}
   -\frac{1}{K'(\alpha)} = \frac{\alpha}{1-\beta} + o(\alpha) .
    \label{1_Kprim_tau<1}
\end{equation}
From the first equation we see that 
$\mu = K(\alpha) \rightarrow \infty$ when $\alpha\rightarrow 0^+$. In other words, the critical
point escapes to infinity. Extracting $\alpha$ from
(\ref{K_tau<1}) and substituting it into (\ref{1_Kprim_tau<1})
we see that $\psi'(\mu)$ (\ref{psi_prim_par}) falls of exponentially for $\mu\rightarrow +\infty$
\begin{equation}
    \psi'(\mu) = - \frac{\left(\Gamma(1-\beta)\right)^{1/(1-\beta)}}{1-\beta} e^{-\frac{\mu}{1-\beta}} + 
    o\left(e^{-\frac{\mu}{1-\beta}}\right) .
    \label{psi_prim_exponential}
\end{equation}
In particular, for $\beta=0$, we see that $\psi'(\mu)$ falls off exponentially for 
$\mu\rightarrow \infty$ as $\psi'(\mu) \approx -e^{-\mu}$. This is consistent with
the exact solution for $\beta=0$. For $\beta=0$ the grand-canonical partition function
(\ref{ZSmu}) can be calculated to give $Z_{S,\mu} = e^{-\mu} (1+e^{-\mu})^{S-1}$ from
$Z_{S,N}$ (\ref{Ztrivial}). In the thermodynamic limit we therefore find: 
$\psi(\mu) = -\log(1 + e^{-\mu})$ and $\psi'(\mu) = -e^{-\mu}/(1+e^{-\mu}) \approx -e^{-\mu}$.


\begin{thebibliography}{99}

\bibitem{bbj} P. Bialas, Z. Burda, and D. Johnston,
Nucl.Phys. B 493, 505 (1997) [arXiv:cond-mat/9609264].

\bibitem{dgc}
J.M. Drouffe, C. Godrèche, and F. Camia,
J. Phys. A 31, L19 (1998) [arXiv:cond-mat/9708010].

\bibitem{bbj2} P. Bialas, Z. Burda, and D. Johnston,
Nucl. Phys. B 542, 413  (1999) [arXiv:gr-qc/9808011].

\bibitem{s} F. Spitzer,  Advances in Math. 5, 246 (1970).

\bibitem{e} M.R. Evans, Braz. J. Phys. 30, 42 (2000)
[arXiv:cond-mat/0007293].

\bibitem{jmp} I. Jeon, P. March, and B. Pittel,
Ann. Probab. 28, 1162 (2000).

\bibitem{eh} M.R. Evans and T. Hanney, J. Phys. A 38, R195
(2005) [arXiv:cond-mat/0501338].

\bibitem{gl} C. Godrèche and J.M. Luck, 
J. Phys. A 38, 7215 (2005) [arXiv:cond-mat/0505640].

\bibitem{kmh} J. Kaupuzs, R. Mahnke, and R. J. Harris, Phys. Rev. E 72, 056125 (2005) [arXiv:cond-mat/0504676].

\bibitem{g} C. Godrèche, Lect. Notes Phys. 716, 261 (2007) 
[arXiv:cond-mat/0604276].

\bibitem{wbbj} B. Waclaw, L. Bogacz, Z. Burda, and
W. Janke,  Phys. Rev. E 76, 046114 (2007) 
[arXiv:cond-mat/0703243].

\bibitem{mez} S. N. Majumdar, M. R. Evans, and R. K. P. Zia
Phys. Rev. Lett. 94, 180601 (2005) [arXiv:cond-mat/0501055].

\bibitem{emz} M. R. Evans, S. N. Majumdar, and R. K. P. Zia,
 Journal of Statistical Physics 123, 357 (2006)
[arXiv:cond-mat/0510512].

\bibitem{emz2} M. R. Evans, S. N. Majumdar, and R. K. P. Zia,
 J. Phys. A: Math. Gen, 39, 4859 (2006) [arXiv:cond-mat/0602564].

\bibitem{bb}  P. Bialas and Z. Burda, 
Phys. Lett. B 384, 75 (1996) [arXiv:hep-lat/9605020].

\bibitem{j} S. Janson, Prob. Surveys 9, 103 (2012).

\bibitem{bb2}  P. Bialas and Z. Burda, Phys. Lett. B 416, 281 (1998) [arXiv:hep-lat/9707028]. 

\bibitem{bbw} L. Bogacz, Z. Burda, and B. Waclaw, Phys. Rev. D 86, 104015 (2012) [arXiv:1204.1356].

\bibitem{bjjknpz} Z. Burda, D. Johnston, J. Jurkiewicz, M. Kamiński, M. A. Nowak, G. Papp and I. Zahed,
Phys. Rev. E 65, 026102 (2002) [arXiv:cond-mat/0101068].



\bibitem{maac}  O. Mazzarisi, A. de Azevedo-Lopes, J. J. Arenzon, and F. Corberi, Phys. Rev. Lett. 127, 128301 (2021) [arXiv:2103.09143].



\bibitem{klmst} Y. Kafri, E. Levine, D. Mukamel, G. M. Sch\"{u}tz, and J. T\"{o}r\"{o}k, Phys. Rev. Lett. 89, 035702 (2002)
[arXiv:cond-mat/0204319].

\bibitem{elmm} M. R. Evans, E. Levine, P. K. Mohanty, and D. Mukamel, Eur. Phys. J. B 41, 223 (2004)
[arXiv:cond-mat/0405049]. 

\bibitem{ahe} A. G. Angel, T. Hanney, and M. R. Evans, Phys. Rev. E 73, 016105 (2006) [arXiv:cond-mat/0509238].

\bibitem{bbw2} P. Bialas, Z. Burda, and B. Waclaw, AIP Conf. Proc. 776, 14 (2005) [arXiv:cond-mat/0503548].

\bibitem{vH} L. Van Hove, Phys. Rev. 89, 1189 (1953).

\bibitem{w} J. G. Wendel, Math. Scand. 14, 21 (1964). 
\bibitem{g1} C. Godrèche, J. Phys. A 50, 195003 (2017)
[arXiv:1611.01434].
\bibitem{g2} C. Godrèche, J. Phys. A 54, 038001 (2021) 
[arXiv:1909.11540].

\bibitem{seva} 
D. Denisov, A. B. Dieker, and V. Shneer,
The Annals of Probability 36, Issue 5, 1946 (2008) 
[arXiv:math/0703265].

\bibitem{gss} S. Grosskinsky, G.M. Sch\"{u}tz, and H. Spohn, J. Stat. Phys. 113, 389 (2003) [arXiv:cond-mat/0302079].

\bibitem{fls} P. A. Ferrari, C. Landim, and V. V. Sisko,
J. Stat. Phys. 128, 1153 (2007) 
[arXiv:math/0612856].

\bibitem{cg1} P. Chleboun and S. Grosskinsky, J. Stat. Phys. 140,
846 (2010) [arXiv:1004.0408].

\bibitem{agl} I. Armendariz, S. Grosskinsky, and M. Loulakis, 
Stoch. Proc. Appl. 123, 3466 (2013) [arXiv:0912.1793].

\bibitem{cg2} P. Chleboun and S. Grosskinsky, 
J. Stat. Phys. 154, 432 (2014) [arXiv:1306.3587].

\bibitem{jcg} W. Jatuviriyapornchai, P. Chleboun, and S. Grosskinsky, 
J. Stat. Phys. 178, 682 (2020) [arXiv:1907.12166].



\bibitem{g3} C. Godrèche, J. Stat. Phys. 182, 13 (2021) 
[arXiv:2006.04076].

\bibitem{mez1} M. R. Evans, S. N. Majumdar, and R. K. P. Zia,
J. Stat. Phys. 123, 357 (2006)
[arXiv:cond-mat/0510512].

\bibitem{mez2}M. R. Evans, S. N. Majumdar, and R. K. P. Zia,
Phys. Rev. Lett. 94, 180601 (2005)
[arXiv:cond-mat/0501055].


\bibitem{gk} B.V. Gnedenko and A. N. Kolmogorov, {\em Limit distributions for sums of independent random variables: Revised Edition}, (Addison-Wesley, Cambridge, 1968, ISBN 978-1-684-22579-8).

\bibitem{gs} B. Gaveau and L. Schulman, J. Phys. A20,  2865 (1987).

\bibitem{rs} G. Roepstor and  L. Schulman, J. Stat. Phys 34,  35 (1984).

\bibitem{as} M. Abramowitz and I.A. Stegun, {\em Handbook of Mathematical Functions with Formulas, Graphs, and Mathematical Tables}, 
(Dover Publications, New York, 1972, ISBN 978-0-486-61272-0).


\bibitem{bbjr} P. Bialas, Z. Burda, and D. Johnston,
The R\'enyi entropy of zeta-urns [arXiv:2307.14472].

\bibitem{wbe} A.J. Wood, R. A. Blythe, and M. R. Evans,
J. Phys. A: Math. Theor. 50, 475005 (2017) [arXiv:1708.00303]. 


\bibitem{bbjz} P. Bialas, Z. Burda, and D. Johnston, Partition function zeros of  zeta-urns, in preparation.


\end{thebibliography}
\end{document}